\documentstyle[a4,12pt]{article}
\begin{document}
\parskip=5pt plus 1pt minus 1pt

\begin{flushright}
{\bf MPI--PhT/95--08}\\
{\today}
\end{flushright}

\begin{center}
{\Large\bf A Symmetry Pattern of Maximal $CP$ Violation \\
and a Determination of the Unitarity Triangle}$^{~*}$
\end{center}

\vspace{0.3cm}

\begin{center}
{\bf Harald Fritzsch} \\
{\sl Sektion Physik, Theoretische Physik, Universit$\ddot{a}$t
M$\ddot{u}$nchen,}\\
{\sl Theresienstrasse 37, D-80333 M$\ddot{u}$nchen, Germany} \\
{\sl and} \\
{\sl Max-Plank-Institut f$\sl\ddot{u}$r Physik ---
Werner-Heisenberg-Institut,}\\
{\sl F$\sl\ddot{o}$hringer Ring 6, D-80805 M$\sl\ddot{u}$nchen, Germany}
\end{center}

\begin{center}
{\bf Zhi-zhong Xing} \\
{\sl Sektion Physik, Theoretische Physik, Universit$\ddot{a}$t
M$\ddot{u}$nchen,} \\
{\sl Theresienstrasse 37, D-80333 M$\ddot{u}$nchen, Germany}
\end{center}
\vspace{1.cm}

\begin{abstract}

Within a specific texture of the quark mass matrix the notion of
a maximal violation of the $CP$ symmetry can be defined. The experimental
constraints
from weak decays imply that in reality one is close to the case of maximal $CP$
violation, which vanishes as the mass of the $u$-quark approaches zero.
The unitarity triangle of the Cabibbo-Kobayashi-Maskawa matrix elements
is determined. It is related to the triangle relating the Cabibbo angle to the
quark mass ratios in the complex plane. The angle $\alpha$ describing the $CP$
violation in the decay $B_{d}\rightarrow \pi^{+}\pi^{-}$ is close to $90^{0}$.

\end{abstract}

\vspace{4cm}

\vspace{1cm}

\begin{flushleft}
\hspace{0cm}{--------------------------------------------} \\
$^{*}$ {\footnotesize Supported in part by DFG-contract 412/22-1, EEC-contract
SC1-CT91-0729, and EEC-contract
CHRX-CT94-0579 (DG 12 COMA)} \\
\end{flushleft}

\newpage

In the standard electroweak model $CP$ violation arises due to the presence
of a complex phase parameter in the Cabibbo-Kobayashi-Maskawa (CKM) mixing
matrix [1,2]. It is well-known that a complex phase enters the CKM matrix
in general after the diagonalization of the quark mass matrices [2],
provided there are at least three families of quarks. Furthermore it is
known that $CP$ violation is absent if the CKM matrix has at least one
vanishing
matrix elements [3]. In case of two families the quark mixing matrix can
always be made real by a suitable redefinition of the phases of the quark
fields.

In the standard electroweak model the masses of the quarks and the weak mixing
angles including the complex phase parameter causing $CP$ violation enter as
free parameters. Further insights into the dynamics which determines the masses
and angles imply steps beyond the electroweak standard model. In the absence
of specific hints towards the underlying dynamics one is invited to consider
specific symmetry schemes which could reduce the number of free parameters in
the fermion sector and eventually provide hints towards further dynamical
details of the mass generation mechanism [4].

In this paper we shall consider a specific hierarchical scheme for the mass
generation of the first family. It predicts the CKM matrix mixing parameters
involving the $u$ and $d$ quarks as well as the $CP$ violating phase parameter.

Our starting point is to consider the limit in which only the masses of the
third
family are present: $m_{u}=m_{c}=m_{d}=m_{s}=0$, $m_{t}\neq 0$ and
$m_{b}\neq 0$. As discussed in ref. [5], this limit can be achieved in a
natural
way such that the ($t$, $b$) system remains unmixed, i.e., the CKM mixing
matrix is
unity.

In a further step (denoted below as step I), the masses of the members of the
second family are introduced, but $m_{u}$ and $m_{d}$ stay massless. As
pointed out in ref. [5], this limit can be achieved in a natural way only
if the first family remains unmixed in the limit, i.e., the up and down
mass matrices will take the following forms simultaneously, which we
arrange to be real and symmetric by a suitable transformation of the
right-handed quark fields:
\begin{equation}
M \; = \; \left (
\begin{array}{ccc}
0  & 0  & 0 \\
0  & C  & B \\
0  & B  & A
\end{array}
\right ) \; .
\end{equation}
After diagonalization of the mass matrix (1) the positive eigenstates are
$\tilde{m_{c}}$, $\tilde{m_{t}}$; $\tilde{m_{s}}$, $\tilde{m_{b}}$
respectively.
We denote the masses in the limit $m_{u}$, $m_{d}\rightarrow 0$ by
$\tilde{m_{q}}$ in order to distinguish them from the actual quark masses,
which differ (except for $m_{u,d}$) from $\tilde{m_{q}}$ by a small correction.
The flavour mixing matrix $\tilde{V}$ takes the form
\begin{equation}
\tilde{V} \; =\; \left (
\begin{array}{ccc}
1       &       0       &       0 \\
0       & \tilde{V}_{cs}        & \tilde{V}_{cb} \\
0       & \tilde{V}_{ts}        & \tilde{V}_{tb}
\end{array}
\right ) \; = \left (
\begin{array}{ccc}
1       &       0       &       0 \\
0       & \tilde{c}_{23}        & \tilde{s}_{23} \\
0       & - \tilde{s}_{23}      & \tilde{c}_{23}
\end{array}
\right ) \; ,
\end{equation}
where $\tilde{c}_{23}=\cos\tilde{\theta}_{23}$ and $\tilde{s}_{23}
=\sin\tilde{\theta}_{23}$ ($\tilde{\theta}_{23}$ is the mixing angle).
No complex phase is obtained in the limit considered here. Note that complex
phases
in the off-diagonal matrix elements multiplying the $\bar{s}b$, $\bar{b}s$ or
$\bar{c}t$, $\bar{t}c$ terms can be eliminated by a phase rotation of the $s$
or
$c$ fields. $CP$ violation is absent. To lowest order approximation
we obtain
\begin{equation}
\tilde{V}_{cb}\; \approx \; - \tilde{V}_{ts}\; \approx \; \frac{B_{\rm
d}}{A_{\rm d}}
-\frac{B_{\rm u}}{A_{\rm u}} \; .
\end{equation}
The mixing angle
$\tilde{s}_{23}=\tilde{V}_{cb}$ might be related to the ratios
$\tilde{m}_{s}/\tilde{m}_{b}$ and $\tilde{m}_{c}/\tilde{m}_{t}$, if
a special texture for the mass matrices exists. For example, in the approach
based on the so-called ``democratic symmetry'' one finds in lowest order of the
symmetry breaking [6]:
\begin{equation}
\tilde{V}_{cb} \; \approx \; \frac{1}{\sqrt{2}} \left
(\frac{\tilde{m}_{s}}{\tilde{m}_{b}}
+ \frac{\tilde{m}_{c}}{\tilde{m}_{t}}\right ) \; .
\end{equation}

The next step (step II) is to introduce the masses of the $d$ quark, but
keeping $m_{u}$ massless. We regard this sequence as useful due to the
fact that the mass ratios $m_{u}/m_{c}$ and $m_{u}/m_{t}$ are about one order
of magnitude smaller than the ratios $m_{d}/m_{s}$ and $m_{d}/m_{b}$.
It is well-known that the observed magnitude of the mixing between the
first and the second family can be reproduced well by a specific texture
of the mass matrix [7,8]. We shall incorporate this here and take the following
ansatz for the mass matrix of the down-type quarks:
\begin{equation}
M_{\rm d} \; = \; \left (
\begin{array}{ccc}
0  & D_{\rm d}  & 0 \\
D^{*}_{\rm d}  & C_{\rm d}  & B_{\rm d} \\
0  & B_{\rm d}  & A_{\rm d}
\end{array}
\right ) \; .
\end{equation}
Here $D_{\rm d}$ is a complex parameter: $D_{\rm d}=|D_{\rm d}|e^{i\sigma_{\rm
d}}$.
At this stage the mass matrix of the up-type quarks remains in the form (1).
The CKM matrix elements $V_{us}$, $V_{cd}$ and the ratios $V_{ub}/V_{cb}$,
$V_{td}/V_{ts}$
can be calculated in this limit. One finds in lowest order:
\begin{equation}
V_{us} \; \approx \; \sqrt{\frac{m_{d}}{m_{s}}} \; , \;\;\;\;\;\;\;
V_{cd} \; \approx \; - \sqrt{\frac{m_{d}}{m_{s}}}e^{-i\sigma_{\rm d}} \; ,
\;\;\;\;\;\;\;
\frac{V_{ub}}{V_{cb}} \; \approx \; 0 \; , \;\;\;\;\;\; \frac{V_{td}}{V_{ts}}
\; \approx
\; - \sqrt{\frac{m_{d}}{m_{s}}} \; .
\end{equation}

An interesting implication of the ansatz (5) is the vanishing of $CP$
violation.
Although the mass matrix (5) contains a complex phase parameter, it can be
rotated away due to the fact that $m_{u}$ is still massless (a phase rotation
of the $u$-field does not lead to any observable consequences). The vanishing
of $CP$ violation can be seen explicitly as follows. Considering the two
hermitian mass matrices
$M_{\rm u}$ and $M_{\rm d}$ in general, one can define the commutator
\begin{equation}
\left [ M_{\rm u}, M_{\rm d} \right ] \; = \; i{\cal C} \; .
\end{equation}
Its determinant ${\rm Det}~{\cal C}$ is a rephasing invariant
measure of $CP$ violation [9]. With the forms of $M_{\rm u}$ and $M_{\rm d}$
given in (1) and (5) respectively, we find
\small
\begin{equation}
{\rm Det}~{\cal C} \; = \; -i \left |
\begin{array}{ccc}
0       & -C_{\rm u}D_{\rm d}   & -B_{\rm u}D_{\rm d} \\
C_{\rm u}D^{*}_{\rm d}  & 0     & -(A_{\rm u}-C_{\rm u})B_{\rm d}+(A_{\rm
d}-C_{\rm d})B_{\rm u} \\
B_{\rm u}D^{*}_{\rm d}  & (A_{\rm u}-C_{\rm u})B_{\rm d}-(A_{\rm d}-C_{\rm
d})B_{\rm u} & 0
\end{array} \right | \; = \; 0 \; .
\end{equation}
\normalsize
The vanishing of $CP$ violation in the limit $m_{u}\rightarrow 0$ in our
approach
is an interesting phenomenon, since it is the same limit in which the
``strong''
$CP$ violation induced by instanton effects of QCD is absent [10]. Whether this
link
between ``strong'' and ``weak'' $CP$ violation could offer a solution of the
``strong'' $CP$ problem remains an open issue at the moment. Nevertheless it is
an interesting feature of our approach that $CP$ violation and the mass of the
$u$ quark
are intrinsically linked to each other.

The final step (step III) is to introduce the mass of the $u$ quark. The mass
matrix
$M_{\rm u}$ takes the form:
\begin{equation}
M_{\rm u} \; = \; \left (
\begin{array}{ccc}
0  & D_{\rm u}  & 0 \\
D^{*}_{\rm u}  & C_{\rm u}  & B_{\rm u} \\
0  & B_{\rm u}  & A_{\rm u}
\end{array}
\right ) \; .
\end{equation}
Once the mixing term $D_{\rm u}=|D_{\rm u}|e^{i\sigma_{\rm u}}$ for the
$u$-quark is introduced, $CP$ violation
appears. For the determinant of the commutator (6) we find:
\begin{equation}
{\rm Det}~{\cal C} \; = \; T_{1}\sin \Delta\sigma +
T_{2} \sin 2\Delta\sigma \; ,
\end{equation}
where $\Delta\sigma\equiv \sigma_{\rm u}-\sigma_{\rm d}$, and
\begin{equation}
\begin{array}{lll}
T_{1} & = & 2 |D_{\rm u}D_{\rm d}| \left [ (A_{\rm u}B_{\rm d}-B_{\rm u}A_{\rm
d})^{2}
-|D_{\rm u}|^{2}B_{\rm d}^{2}-B_{\rm u}^{2}|D_{\rm d}|^{2} \right . \\
&  &    \left . -(A_{\rm u}B_{\rm d}-B_{\rm u}A_{\rm d})(C_{\rm u}B_{\rm
d}-B_{\rm u}C_{\rm d})\right ] \; , \\
T_{2} & = & 2B_{\rm u}B_{\rm d}|D_{\rm u}D_{\rm d}|^{2} \; .
\end{array}
\end{equation}
The phase difference $\Delta\sigma$ determines the strength of $CP$ violation.
The vanishing of $CP$ violation in the limit $m_{u}\rightarrow 0$ can be seen
directly, since both $T_{1}$ and $T_{2}$ vanish as $m_{u}\rightarrow 0$.

The diagonalization of the mass matrices $M_{\rm d}$ and $M_{\rm u}$ leads to
the
eigenvalues $m_{i}$ ($i=u,d,...$). Note that $m_{u}$ and $m_{d}$ appear to be
negative.
By a suitable $\gamma_{5}$-transformation of the quark fields one can arrange
them to be
positive. Collecting the lowest order terms in the CKM matrix, one obtains:
\begin{equation}
V_{us} \; \approx \; \sqrt{\frac{m_{d}}{m_{s}}}
-\sqrt{\frac{m_{u}}{m_{c}}}e^{i\Delta\sigma} \; , \;\;\;\;\;\;\;
V_{cd} \; \approx \; \sqrt{\frac{m_{u}}{m_{c}}}
-\sqrt{\frac{m_{d}}{m_{s}}}e^{i\Delta\sigma} \;
\end{equation}
and
\begin{equation}
\frac{V_{ub}}{V_{cb}} \; \approx \; - \sqrt{\frac{m_{u}}{m_{c}}} \; ,
\;\;\;\;\;\;\;\;\; \frac{V_{td}}{V_{ts}} \; \approx \; -
\sqrt{\frac{m_{d}}{m_{s}}} \; .
\end{equation}
The relations for $V_{us}$ and $V_{cd}$ were obtained previously by one of us
(H.F.) [8].
However it was not noted that the relative phase between the two mass ratios
might be of special relevance for $CP$ violation. A related discussion can
be found in ref. [11].

According to eq. (10) the strength of $CP$ violation depends on the phase
difference
$\Delta\sigma$. If we keep the moduli of the parameters
$D_{\rm u}$ and $D_{\rm d}$ constant, but vary the phase from zero to $90^{0}$,
the
strength of $CP$ violation varies from zero to a maximal value given by eq.
(10).
The latter  is determined by the ratio $T_{2}/T_{1}$ . In a good approximation
one finds
for this ratio:
\begin{equation}
\left |\frac{T_{2}}{T_{1}}\right | \; \approx \;
\sqrt{\frac{m_{u}m_{d}}{m_{c}m_{s}}}
\cdot\frac{m_{c}m_{s}}{m^{2}_{t}m^{2}_{b}}\cdot\frac{|B_{\rm u}B_{\rm
d}|}{|V_{cb}|^{2}} \; .
\end{equation}
The mixing parameter $B_{\rm u}$ and $B_{\rm d}$, which determine the mixing
between
the third and the second families, must be small in comparison to $m_{t}$ and
$m_{b}$
respectively, at most of the order of $\sqrt{m_{c}m_{t}}$ or
$\sqrt{m_{s}m_{b}}$
respectively. A generous bound on the ratio is $|T_{2}/T_{1}|< 2\times
10^{-5}$. Thus the coefficient
of the second term in eq. (10) is very small in comparison to the coefficient
of the
first term. We conclude that in a very good approximation (better than
$0.002^{0}$)
$CP$ violation is maximal for $\Delta\sigma=90^{0}$. In this case the element
$D_{\rm u}$ would be purely imaginary, if we set the phase of the matrix
element
$D_{\rm d}$ to be zero (this can always be arranged).

In our approach the $CP$-violating phase also enters in the expressions for
$V_{us}$
and $V_{cd}$ (Cabibbo angle). As discussed already in refs. [4,8], the Cabibbo
angle
is fixed by the difference of $\sqrt{m_{d}/m_{s}}$ and $\sqrt{m_{u}/m_{c}}$
$\times$ phase factor.
The second term contributes a small correction (of order 0.06) to the leading
term, which according to the mass ratios given in ref. [12] is allowed to vary
between
0.20 and 0.24. For our subsequent discussion we shall use $0.218\leq
|V_{us}|\leq 0.224$ [12].
If the phase parameter multiplying $\sqrt{m_{u}/m_{c}}$ were zero or $\pm
180^{0}$
(i.e. either the difference or sum of the two real terms would enter), the
observed magnitude
of the Cabibbo angle could not be reproduced. Thus a phase is needed. We find
within
our approach purely on phenomenological grounds (not related to the observed
$CP$ violation
in $K$-meson decays) that $CP$ violation must be present if
we request consistency between observation and our result given in eq. (12).

An excellent description of the magnitude of $V_{us}$ is obtained for a phase
angle of $90^{0}$.
In this case one finds:
\begin{equation}
|V_{us}|^{2} \; \approx \; \left (1-\frac{m_{d}}{m_{s}}\right )
\left (\frac{m_{d}}{m_{s}}+\frac{m_{u}}{m_{c}}\right ) \; ,
\end{equation}
where approximations are made for $V_{us}$ to a better degree of accuracy than
that in eq. (12).
Using $|V_{us}|$ = $0.218\ldots 0.224$ and $m_{u}/m_{c}$ = $0.0028\ldots
0.0048$ we obtain
$m_{d}/m_{s}$ $\approx$ $0.045\ldots 0.05$. This corresponds to $m_{s}/m_{d}$
$\approx$ $20\ldots 22$,
which is entirely consistent with the determination of $m_{s}/m_{d}$, based on
chiral
perturbation theory [13]: $m_{s}/m_{d}$ = $17\ldots 25$. This example shows
that the phase angle
must be in the vicinity of $90^{0}$. The general situation is described in
${\rm fig.~ 1}$.
Fixing $m_{u}/m_{c}$ to its central value and varying $m_{d}/m_{s}$ and
$|V_{us}|$ throughout the
allowed ranges, we find $\Delta \sigma$ $\approx$ $66^{0}\ldots 110^{0}$.

The case $\Delta \sigma=90^{0}$, favoured by our analysis, deserves a special
attention. It implies that in the sequence of steps discussed above the term
$D_{\rm u}$ generating the mass of the $u$-quark is purely imaginary, and hence
$CP$ violation is maximal. It is interesting to observe that nature seems to
prefer
this case. A purely imaginary term $D_{\rm u}$ implies that the algebraic
structure
of the quark mass matrix is particularly simple. Its consequences need to be
investigated further and might lead the way towards a deeper dynamical
understanding of
the pattern of masses and of $CP$ violation.

Finally we explore the consequences of our approach for the unitarity triangle,
i.e., the triangle formed by the CKM matrix elements $V_{ud}V^{*}_{ub}$,
$V_{cd}V^{*}_{cb}$
and $V_{td}V^{*}_{tb}$ in the complex plane (see fig. 2(a)):
\begin{equation}
V_{ud}V^{*}_{ub} + V_{cd}V^{*}_{cb} + V_{td}V^{*}_{tb} = 0 \; .
\end{equation}
Here we use the definitions of the angles $\alpha$, $\beta$ and $\gamma$ as
given in ref. [12].
For $\Delta\sigma=90^{0}$ we obtain:
\begin{equation}
\alpha \; \approx \; 90^{0} \; , \;\;\;\;\;\;\;
\beta \; \approx \; \arctan\sqrt{\frac{m_{u}}{m_{c}}\cdot
\frac{m_{s}}{m_{d}}} \; , \;\;\;\;\;\;\;
\gamma \; \approx \; 90^{0}-\beta \; .
\end{equation}
Thus the unitarity triangle is approximately a rectangular triangle. Using as
input
$m_{u}/m_{c}$ = $0.0028\ldots 0.0048$ and $m_{s}/m_{d}$ = $20\ldots 22$ as
discussed above,
we find $\beta$ $\approx$ $13^{0}\ldots 18^{0}$, $\gamma$ $\approx$
$72^{0}\ldots 77^{0}$,
and $\sin 2\beta$ $\approx$ $\sin 2\gamma$ $\approx$ $0.45\ldots 0.59$.
These values are consistent with the existing experimental constraints [14].

We note that the unitarity triangle as described above and the triangle formed
in the complex
plane by $V_{cd}$, $\sqrt{m_{u}/m_{c}}$ and $\sqrt{m_{d}/m_{s}}e^{i\Delta
\sigma}$ (see
eq. (12) and fig. 2(b)), to which we shall refer as the Cabibbo triangle, are
in lowest order
related by a similarity transformation. This can be seen as follows. With the
help of eq. (13)
as well as the lowest order result
\begin{equation}
V_{ud}\; \approx \; 1 \; , \;\;\;\;\;\;
V_{cs}\; \approx \; V_{tb} \; \approx  \; e^{i\Delta \sigma} \; , \;\;\;\;\;\;
V_{ts}e^{-i\Delta \sigma} \; \approx \; -V^{*}_{cb}e^{i\Delta \sigma} \; ,
\end{equation}
we transform the unitarity relation (16) into
\begin{equation}
-\sqrt{\frac{m_{u}}{m_{c}}}V_{cb}^{*} + V_{cd}V^{*}_{cb} +
\sqrt{\frac{m_{d}}{m_{s}}}V^{*}_{cb}e^{i\Delta \sigma}
\; \approx \; 0 \; .
\end{equation}
After removing an overall factor $V_{cb}^{*}$ on the left-handed side of eq.
(19), one finds:
\begin{equation}
V_{cd} \; \approx \; \sqrt{\frac{m_{u}}{m_{c}}} -
\sqrt{\frac{m_{d}}{m_{s}}}e^{i\Delta \sigma} \; .
\end{equation}
This is just our result (12) determining the Cabibbo triangle. The Cabibbo
triangle and the
unitarity triangle are identical apart from the overall scale.
It is interesting to note that the relation (20) or (12), which is not
directly related to $CP$ violation,  determines the $CP$ violating phase
$\Delta\sigma$ ($\approx \alpha$).

Finally we add a comment about the phase $\delta $ used in the standard
parametrization of the CKM matrix given in ref.\ (12). Here the $CP$
violating phase enters primarily in the $V_{ub}$ matrix element:
$V_{ub} = s_{13} \, e^{-i \delta }$. Inspecting the unitarity triangle, one
finds after taking into account that $s_{12} V^*_{cb}$ ($s_{12}$:
Cabibbo angle) is real: $\delta = \gamma$, i.e.\ $\delta \approx 90^{\circ }
- arc tan \sqrt{\frac{m_u \cdot m_s}{m_c \cdot m_d}} \approx 72^{\circ }
\ldots 77^{\circ }$. Note that maximal $CP$ violation as defined by us
does not imply a maximal phase $\delta $.

In this paper we have shown that a simple pattern for the generation of masses
for the first family of leptons and quarks leads to an interesting and
predictive pattern for the violation of $CP$ symmetry. The observed magnitude
of the Cabibbo
angle requires $CP$ violation to be maximal or at least near to its maximal
strength.
The ratio $V_{ub}/V_{cb}$ as well as $V_{td}/V_{ts}$ are given by
$\sqrt{m_{u}/m_{c}}$
and $\sqrt{m_{d}/m_{s}}$ respectively. In the case of maximal $CP$ violation
the
unitarity triangle is approximately rectangular ($\alpha\approx 90^{0}$), the
angle $\beta$ can vary in the
range $13^{0}\ldots 18^{0}$ ($\sin 2\beta\approx \sin 2\gamma$ $\approx$
$0.45\ldots 0.59$). It remains
to be seen whether the future experiments, e.g. the measurements of the $CP$
asymmetries
in $B^{0}_{d}$ vs $\bar{B}^{0}_{d}\rightarrow J/\psi K_{S}$ and
$\pi^{+}\pi^{-}$, confirm these values. \\

\underline{Acknowledgements:} $~$ One of us (H.F.) likes to thank A. Ali, C.
Jarlskog,
T. D. Lee, R. Peccei and R. R$\rm\ddot{u}$ckl for useful discussions.
Z.X. is indebted to the Alexander von Humboldt Foundation for its financial
support.


\begin{thebibliography}{99}

\bibitem{1} N. Cabbibo, Phys. Rev. Lett. 10, 531 (1964).

\bibitem{2} M. Kobayashi and T. Maskawa, Prog. Theor. Phys. 49, 652 (1973).

\bibitem{3} H. Fritzsch, Phys. Rev. D 32, 3058 (1985).

\bibitem{4} H. Fritzsch, Nucl. Phys. B 155, 189 (1979); \\
S. Dimopoulos, L. J. Hall and S. Raby, Phys. Rev. Lett. 68, 1984 (1992); \\
P. Ramond, R. G. Roberts and G. G. Ross, Nucl. Phys. B 406, 19 (1993).

\bibitem{5} H. Fritzsch, Phys. Lett. B 184, 391 (1987).

\bibitem{6} H. Fritzsch and D. Holtmannsp$\rm\ddot{o}$tter, Phys. Lett. B 338,
290 (1994).

\bibitem{7} S. Weinberg, Transactions of the New York Academy of Sciences,
Series II, Vol. 38, 185 (1977).

\bibitem{8} H. Fritzsch, Phys. Lett. B 73, 317 (1978).

\bibitem{9} C. Jarlskog, Phys. Rev. Lett. 55, 1039 (1985).

\bibitem{10} R. D. Peccei and H. R. Quinn, Phys. Rev. D 16, 1791 (1977); \\
S. Weinberg, Phys. Rev. Lett. 40, 223 (1978); \\
F. Wilczek, Phys. Rev. Lett. 40, 279 (1978).

\bibitem{11} M. Shin, Phys. Lett. B 145, 285 (1984); \\
M. Gronau, R. Johnson and J. Schechter, Phys. Rev. Lett. 54, 2176 (1985).

\bibitem{12} Particle Data Group, M. Aguilar-Benitez et al., Phys. Rev. D 50,
1173 (1994).

\bibitem{13} J. Gasser and H. Leutwyler, Phys. Rep. 87, 77 (1982); \\
for a recent review of quark mass values, see, e.g., Y. Koide, preprint
US-94-05 (1994).

\bibitem{14} A. Ali and D. London, preprint CERN-TH.7398/94 (to appear in Z.
Phys. C); \\
I. I. Bigi, preprint CERN-TH.7207/94 (invited talk given at the VIII Rencontres
de
Physique de la Vallee d'Aoste, La Thuile, March 1994).
\end{thebibliography}
\end{document}